\documentclass[a4paper,11pt]{article}

\pdfoutput=0 

\usepackage{jheppub} 

\usepackage[T1]{fontenc} 

\def\d{{\rm d}}

\def\beq{\begin{equation}}
\def\eeq{\end{equation}}
\def\bea{\begin{eqnarray}}
\def\eea{\end{eqnarray}}

\title{\boldmath Nuclear parton density modifications from low-mass
 lepton pair production at the LHC}

\author{M.\ Brandt,}
\author{M.\ Klasen,}
\author{F.\ K\"onig}

\affiliation{Institut f\"ur Theoretische Physik, Westf\"alische
 Wilhelms-Universit\"at M\"unster, \\ Wilhelm-Klemm-Stra\ss{}e 9,
 D-48149 M\"unster, Germany}

\emailAdd{matthiasbrandt@uni-muenster.de}
\emailAdd{michael.klasen@uni-muenster.de}
\emailAdd{f.koenig@uni-muenster.de}

\preprint{MS-TP-14-02}

\abstract{In this article, we investigate the potential of low-mass lepton
 pair production in proton-ion collisions at the LHC to constrain nuclear
 modifications of parton densities. Similarly to prompt photon production, 
 the transverse momentum spectrum is shown to be dominated by the QCD Compton
 process, but has virtually no fragmentation or isolation uncertainties.
 Depending on the orientation of the proton and ion beams and on the use of
 central or forward detector components, all interesting regions of nuclear
 effects (shadowing, antishadowing, isospin and EMC effects) can be probed.
 Ratios of cross sections allow to eliminate theoretical scale and bare-proton
 parton density errors as well as many experimental systematic uncertainties.}

\begin{document} 
\maketitle
\flushbottom

\section{Introduction}

An important physics goal of the Large Hadron Collider (LHC) program
at CERN is the study of the quark-gluon plasma, a deconfined state of matter
created at high temperatures in heavy-ion collisions. For an unambiguous
identification of its properties, e.g.\ through thermal radiation of photons with
small transverse momenta \cite{Wilde:2012wc}, a subtraction of cold nuclear effects
is mandatory \cite{Klasen:2013mga}. In particular, the parton density functions
(PDFs) in nucleons are significantly modified when the latter do not collide freely,
but are bound in nuclear states
\cite{Hirai:2007sx,Schienbein:2009kk,Eskola:2009uj,deFlorian:2011fp},
and this must properly be taken into account \cite{Hammon:1998gq}.

Unfortunately, nuclear PDFs are known with much less precision than free proton
PDFs. Both are mainly determined in deep-inelastic scattering (DIS), but for
nuclei only fixed-target data with a very restricted range in Bjorken-$x$ and photon
virtuality $Q^2$ are available. Since DIS is mainly sensitive to the (valence) quark
density, it is supplemented by measurements of Drell-Yan lepton-pair production
(again in fixed-target experiments), which helps to constrain the antiquark densities
and also the quark densities with a different weighting than the DIS data
\cite{Hirai:2007sx,Schienbein:2009kk}. More recently, data on neutral
\cite{Eskola:2009uj,deFlorian:2011fp} and charged \cite{deFlorian:2011fp} pion
production in proton-ion collisions
at BNL's Relativistic Heavy Ion Collider (RHIC) have been added
in order to put first direct constraints on the poorly known gluon density.

It is well known that the gluon density in protons and nuclei can also
be constrained with prompt photon production in hadronic collisions. In particular,
it has been proposed to use RHIC and LHC data on photons with low transverse
momenta produced in $pA$ collisions
to measure the shadowing effects in the low-$x$ region \cite{Arleo:2007js}.
At higher transverse momenta, this process can also become sensitive
to antishadowing and the EMC effect, albeit not the Fermi motion \cite{Arleo:2011gc}.
However, photons coming from pion decay must be eliminated, e.g.\ by applying an
isolation criterion \cite{Catani:2013oma}, and the contribution of quark and gluon
fragmentation to photons must be parametrised \cite{Bourhis:1997yu}, which induces
considerable uncertainties.

In this paper we advocate that these difficulties can be avoided
by the use of virtual photons, which have neither decay nor fragmentation
contributions and need not be isolated from hadronic spectators, since
they give rise to an experimentally well identifiable lepton (electron or
muon) pair. This idea was first introduced for unpolarised \cite{Berger:1998ev}
and polarised \cite{Berger:1999es} $pp$ collisions, recently extended to
weak boson production at very large transverse momenta \cite{Brandt:2013hoa},
and is now applied to $pA$ collisions.
The transverse momentum spectra of massive vector bosons ($J/\Psi$, $\Upsilon$
and $W/Z$) produced in $pA$ collisions have recently been studied elsewhere
\cite{Kang:2012am,Guzey:2012jp}.
At variance with the Drell-Yan process, which is dominated
at low transverse momenta by quark-antiquark annihilation, we show that, in the
region of low virtuality and high transverse momentum of the lepton pair,
its production in $pA$ collisions is dominated, just like the production of prompt photons,
by quark-gluon scattering and is thus
very sensitive to the poorly constrained nuclear modification of the gluon
PDF. We also demonstrate that other uncertainties
like those from scale variations and the free gluon PDFs largely
cancel out in ratios of cross sections and thus do not spoil this sensitivity.

The remainder of the paper is organised as follows: In Sec.\ 2, we
review the current uncertainties on nuclear PDFs, and in Sec.\ 3 we
establish numerically the dominance of quark-gluon scattering. Sec.\
4 contains our main results for low-mass lepton pair production
at the LHC and examines its sensitivity to various nuclear
modification effects. Our conclusions are presented in Sec.\ 5.

\section{Nuclear parton density uncertainties}
\label{sec:2}

First, we review in this section the current level of uncertainty in the distribution
of a parton $i$ in a nucleus $A$, $f_{i/A}(x,\mu_f)$, which depends on the longitudinal
momentum fraction $x$ carried by the parton and the factorisation scale $\mu_f$.
As we will be mostly interested in ratios of cross sections in $pA$ over those in $pp$
collisions, we focus on the uncertainty of the so-called nuclear modification
factor
\bea
 R_{i/A}(x,\mu_f) &=&  {f_{i/A}(x,\mu_f) \over f_{i/p}(x,\mu_f)}
 \label{eq:1}
\eea
rather than the intrinsic uncertainties of the bare proton PDFs $f_{i/p}(x,\mu_f)$,
which should cancel to a large extent in ratios of cross sections. A modification
factor as in Eq.\ (\ref{eq:1}) is also usually employed in global fits of nuclear PDFs
to parametrise nuclear modification effects, the notable exception being the 
one performed by the nCTEQ collaboration \cite{Schienbein:2009kk}. Cross section
ratios will later of course be computed taking the convolution of partonic
cross sections with different PDFs into account.

In this paper, we use the EPS09 nuclear PDFs as our baseline \cite{Eskola:2009uj},
which are based on the factorised ansatz in Eq.\ (\ref{eq:1}) and CTEQ6.1M
\cite{Stump:2003yu} free-proton PDFs. The bound-neutron PDFs are then obtained by
assuming isospin symmetry.
E.g.\ the total up-quark ($u$) distribution per nucleon in a nucleus $A$ with $Z$
protons is
\beq
 f_{u/A}(x,\mu_f) = {Z\over A} [R_{u/A}^v f^v_{u/p} + R^s_{u/A} f^s_{u/p}] +
            {A-Z\over A}  [R_{d/A}^v f^v_{d/p} + R^s_{d/A} f^s_{d/p}],
 \label{eq:2}
\eeq
where $d$ corresponds to the down-quark and
the superscripts $v$ and $s$ refer to valence and sea quark contributions,
respectively. The parametrisation of the nuclear modifications $R_{i/A}(x,\mu_f)$
is performed at the charm quark mass ($m_c$) threshold imposing the
momentum and baryon number sum rules
\beq
 \sum_{i=q,\overline{q},g} \int_0^1 dx \, xf_{i/A}(x,m_c) = 1, \quad \int_0^1 dx \left[ f^v_{u/A}(x,m_c) + f^v_{d/A}(x,m_c) \right] = 3, \label{eq:sumrules}
\eeq
for each nucleus $A$ separately. At higher scales, 
the nuclear PDFs are obtained by solving the DGLAP evolution equations. This
approach results in an excellent fit to the different types of nuclear hard-process
data \cite{Eskola:2009uj}, suggesting that factorisation works well in the energy
range studied and that
the extracted nuclear PDFs are universal in the region $x\geq0.005$, $\mu_f\geq 1.3$ GeV.

In addition to the theoretical uncertainties of the free-proton PDFs, obtained using
the 40 error sets of the CTEQ6.1M parametrisation \cite{Stump:2003yu},
30 error sets are assigned pairwise to the uncorrelated eigendirections
of the 15 parameters fitted to the nuclear collision data sets. A total uncertainty band
at 90\% confidence level is then calculated from the 71 sets defined by
fixing either $R_{i/A}(x,\mu_f)$  to the best fit value and varying the
free-proton PDFs or fixing the latter to its best fit value and varying the former.
These variations then contribute pairwise to the size of the upper and lower errors via
\bea
 \delta^+f&=&\sqrt{\sum_{i}[\max(f_i^{(+)}-f_0,f_i^{(-)}-f_0,0)]^2},\\
 \delta^-f&=&\sqrt{\sum_{i}[\max(f_0-f_i^{(+)},f_0-f_i^{(-)},0)]^2}.
\eea
As the authors acknowledge, this factorised approach represents a
simplification, violating, e.g., in some cases momentum conservation,
so that strictly speaking the free- and bound-proton PDF uncertainty
analyses should not be separated \cite{Eskola:2009uj}.

In order to estimate the bias from different underlying
free-proton PDFs, parametrisations of the nuclear modification,
and fitted nuclear data sets, we also study the best fits of the
HKN07 \cite{Hirai:2007sx}, nCTEQ \cite{Schienbein:2009kk} and DSSZ
\cite{deFlorian:2011fp} collaborations. In particular, the HKN07
nuclear PDFs are based on the MRST1998 free-proton set \cite{Martin:1998sq}
and those of DSSZ on the more recent MSTW2008 set \cite{Martin:2009iq}.
The nCTEQ parametrisation is (so far) the only one that does not rely on
a factorisation into a nuclear modification factor and free-proton PDFs.
Instead it introduces an explicit $A$-dependence in the coefficients
of the $x$-dependent functional form of the PDFs at the starting scale.
Only the technical framework of the CTEQ6M analysis is used here
\cite{Pumplin:2002vw}.

\begin{figure}
 \centering
 \epsfig{file=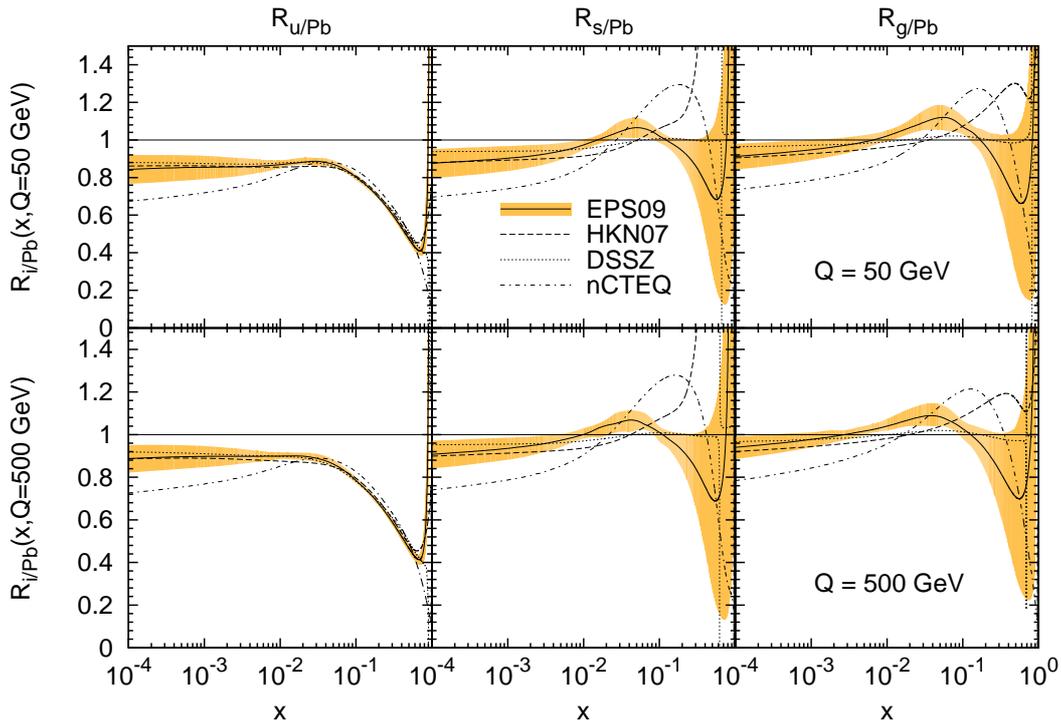,width=0.66\columnwidth,angle=-90}
 \caption{\label{fig:0}Modifications of up-quark (left), strange-quark
 (centre) and gluon (right) PDFs in lead nuclei with respect to bare
 proton PDFs at the factorisation scales 50 GeV (top) and 500 GeV (bottom)
 and as obtained in global fits by four different groups.}
\end{figure}

Since the uncertainties on the free proton PDFs
cancel out to a large extent in ratios of cross sections, we plot in
Fig.\ \ref{fig:0} only the nuclear modification factor defined in
%
%
Eq.\ (\ref{eq:1}) for protons bound in lead ions and as obtained in the
central EPS09 fit (full) with its 30 error sets related to the nuclear
uncertainty (yellow band) \cite{Eskola:2009uj}, in the HKN07
(dashed) \cite{Hirai:2007sx}, nCTEQ (dot-dashed) \cite{Schienbein:2009kk}
and DSSZ (dotted) \cite{deFlorian:2011fp} fits.
Error bands are only shown for the EPS09 analysis. From left to right,
the up-quark, strange-quark and gluon densities are shown separately
for $\mu_f=50$ GeV (top) and 500 GeV (bottom) pertinent to the transverse
momenta of lepton pairs that we will study later.

All four fits agree reasonably well with respect to the nuclear
modification of the up-quark density, which is dominated by valence
quarks and thus mostly modified by strong isospin effects at large $x$.
Qualitatively, the shadowing and antishadowing regions below and
above $x\sim10^{-2}$ as well as the EMC effect and Fermi motion regions
above $x\sim0.1$ and 0.8 are also similar in all distributions.
However, the four fits disagree not only about their sizes, which
vary from 10 to 30 \% in the shadowing and antishadowing regions
and from 0 to 80\% in the region of the EMC
effect, but they also disagree about the relevant $x$-regions.
Only the valence-dominated up-quark distribution at intermediate $x$
is well constrained by the DIS data used in all fits.
Furthermore, the EPS09 parametrisation clearly underestimates the true uncertainty,
as the three other curves often fall outside its band width. This is particularly 
true for the nCTEQ parametrisation, which is obtained without a factorised fit
and has more free parameters than the factorised ansatz of EPS09. A
comparison of Fig.\ \ref{fig:0} with Fig.\ 3 in Ref.\ \cite{Eskola:2009uj}
exhibits that the uncertainties
on these effects decrease when the factorisation scale is increased from
1.3 GeV and 10 GeV as shown there to 50 and 500 GeV as shown here, but only
in the low-$x$ region. This can be traced back to the constraints imposed by
the QCD evolution equations. Note,
however, that the large uncertainties at high $x$ persist, in particular for
the gluon and the sea-dominated strange quark, where DSSZ, e.g.,
predict much smaller nuclear modifications than the other groups.

New experimental information has recently become available
from charged particle production at the LHC, where the CMS collaboration
measure antishadowing enhancements of up to $40\pm20$\% at values of
$x=0.01\,...\,0.05$, i.e.\ much more than predicted by the EPS09 and even nCTEQ
parametrisations \cite{CMS:2013cka}.

\section{Partonic subprocesses in proton-ion collisions}
\label{sec:3}

We now establish in this section which partonic subprocesses
contribute to the production of low-mass lepton pairs at the LHC in
different regions of the transverse momentum ($p_T$) of the pair. This
will later allow us to estimate the sensitivity of the corresponding
experimental measurements on the nuclear modification effects for
quarks and gluons, respectively.
To this end, we compute the transverse momentum spectra of electron-positron
and muon pairs produced in proton-lead collisions at a centre-of-mass energy of
$\sqrt{s}=5.02$ TeV, that has recently been reached at the LHC.

The sensitivities of the ALICE, ATLAS and CMS experiments to continuum production
of lepton pairs in $pp$ collisions have in 2011 allowed to reach transverse momenta
up to 100 GeV and might in the future extend up to 1000 GeV \cite{atlas}.
In these regions, we can rely on a next-to-leading order (NLO) calculation,
and resummation effects at low transverse momenta need not be taken into
account \cite{Berger:1998ev}.
With the ALICE detector, electron and muon pairs are measured in the
$pp$ centre-of-mass frame at rapidities of $|\eta|<0.9$
\cite{Baumann:2012ij} and $2.5<\eta<4$ \cite{Arnaldi:2012bg}, respectively.
In $pA$ collisions, these regions are shifted by a boost of $\Delta\eta=0.465$
in the direction of the proton beam due to the energy asymmetry of the LHC beams
($E_p=4$ TeV, $E_{\rm Pb}=1.58\cdot A$ TeV) \cite{Abelev:2013yxa}.
In order to avoid the $J/\Psi$ and $\Upsilon$ resonance regions, we consider
invariant lepton masses of $2m_\mu<Q<2.5$ GeV and 5 GeV $<Q<6$ GeV.
The inner detectors of the ATLAS \cite{Aad:2012wfa} and CMS \cite{Chatrchyan:2013tia}
experiments allow to measure muons with rapidities $|\eta|<2.4$
in the centre-of-mass frame. The range for electrons is often adjusted
for comparability of the two channels. As a typical invariant mass
range, we will employ here 12 GeV $<Q<25$ GeV. 

\begin{figure}
 \centering
 \epsfig{file=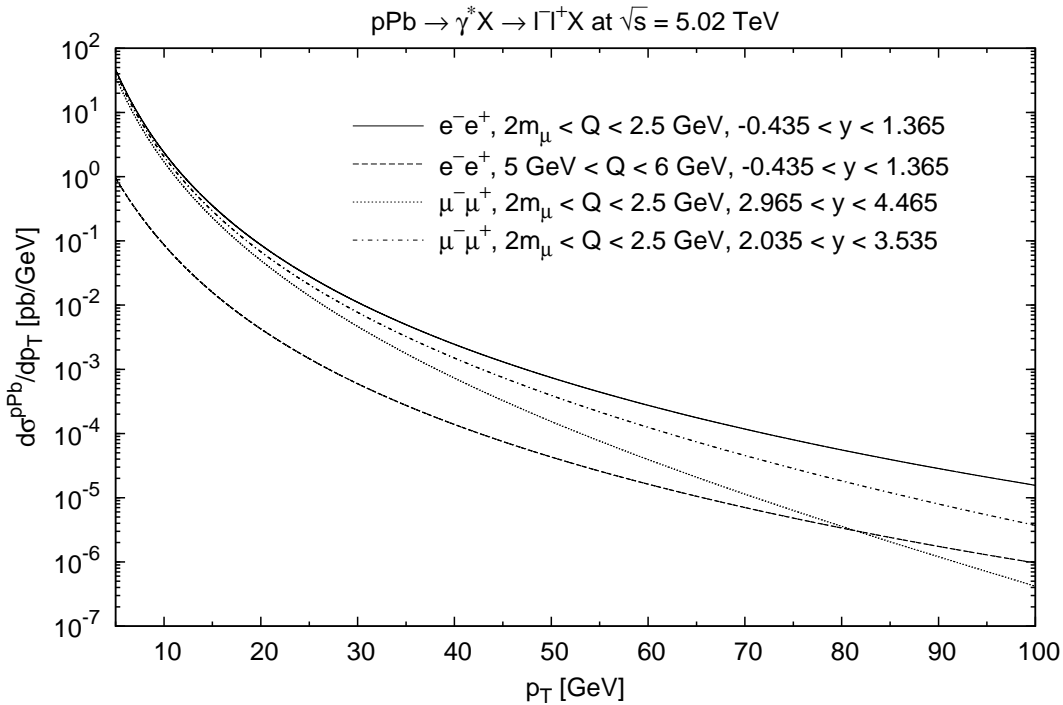,width=0.66\columnwidth,angle=-90}
 \caption{\label{fig:1}Total cross sections for central and forward
 lepton-pair production in proton-lead (full, dashed and dotted lines)
 and lead-proton (dot-dashed line) collisions.}
\end{figure}

Total cross sections for central and forward lepton-pair production
in proton-lead collisions at the LHC with a centre-of-mass energy of
$\sqrt{s}=5.02$ TeV are shown in Fig.\ \ref{fig:1}.
%
%
In the 2013 proton-lead run, a total luminosity of only about 32 nb$^{-1}$
has been delivered, while in the 2013 reference $pp$ run at the same energy
a total luminosity of 5.38 pb$^{-1}$ has been reached. With this luminosity,
centrally produced low-mass electron pairs can be measured with transverse
momenta of up to 50 GeV (full line). For larger invariant masses (dashed line)
or muons in the forward ALICE detector (dotted and dot-dashed lines), the cross sections
are lower by up to an order of magnitude. In principle, much higher luminosities
can of course be achieved at the LHC, as has been demonstrated by the 2012 $pp$
run with an integrated of 23.3 fb$^{-1}$, but for meaningful studies of cold nuclear
effects, it would be necessary to allocate sufficient beam time to proton-lead
collisions.


At leading order (LO) of perturbative QCD, the production of lepton pairs with
finite transverse momentum proceeds through the fusion process $q\bar{q}\to
\gamma^*g$ and through the so-called QCD Compton process $qg\to\gamma^*
q$, where in both cases the virtual photon decays into a lepton pair.
While at low transverse momenta the former process is expected
to dominate and must eventually be matched to the Drell-Yan process
$q\bar{q}\to\gamma^*$ with vanishing transverse momentum, the latter
process is expected to take over as the transverse momentum of
the photon becomes larger, since this also happens for real photons.

\begin{figure}
 \centering
 \epsfig{file=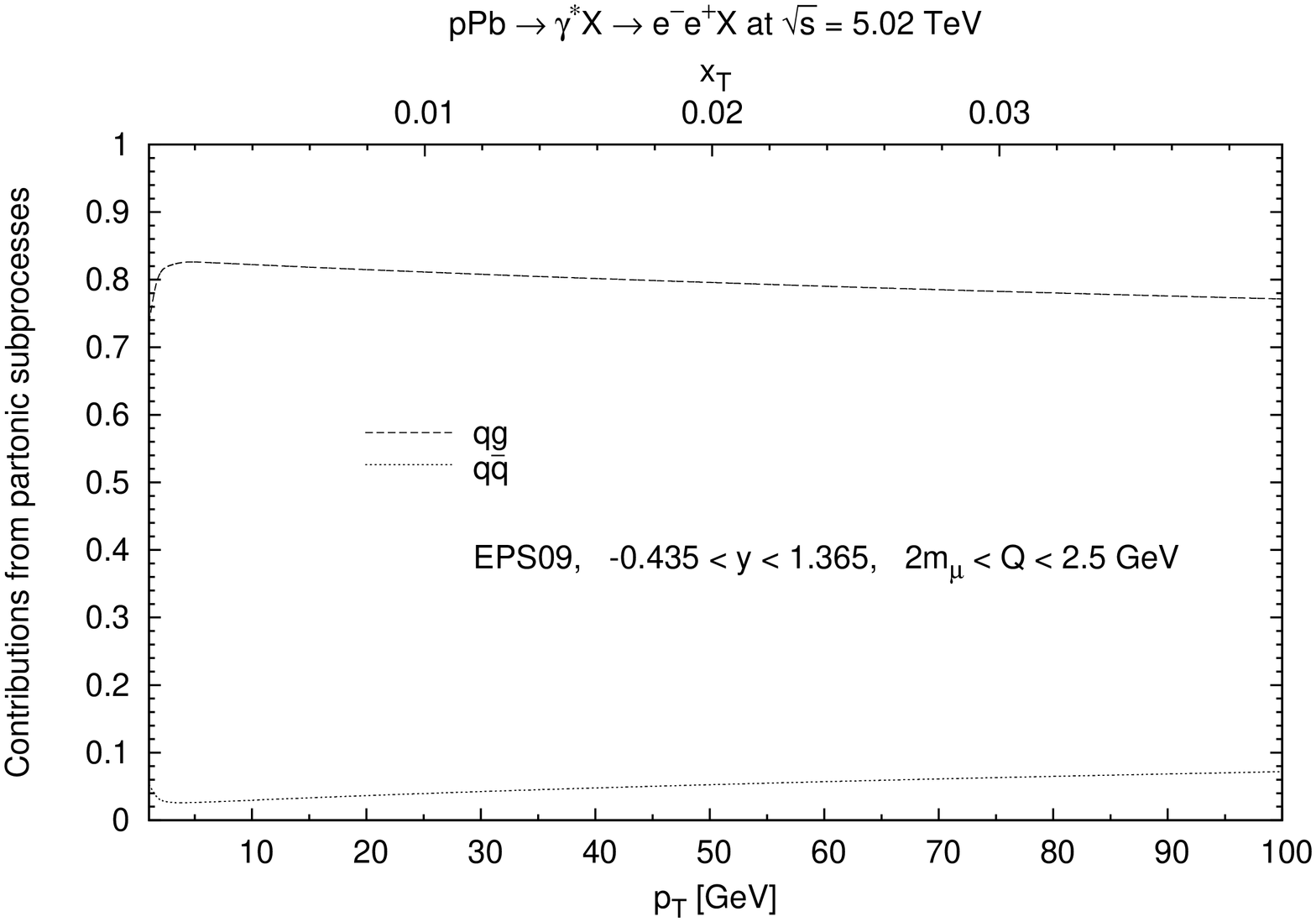,width=0.66\columnwidth,angle=-90}
 \epsfig{file=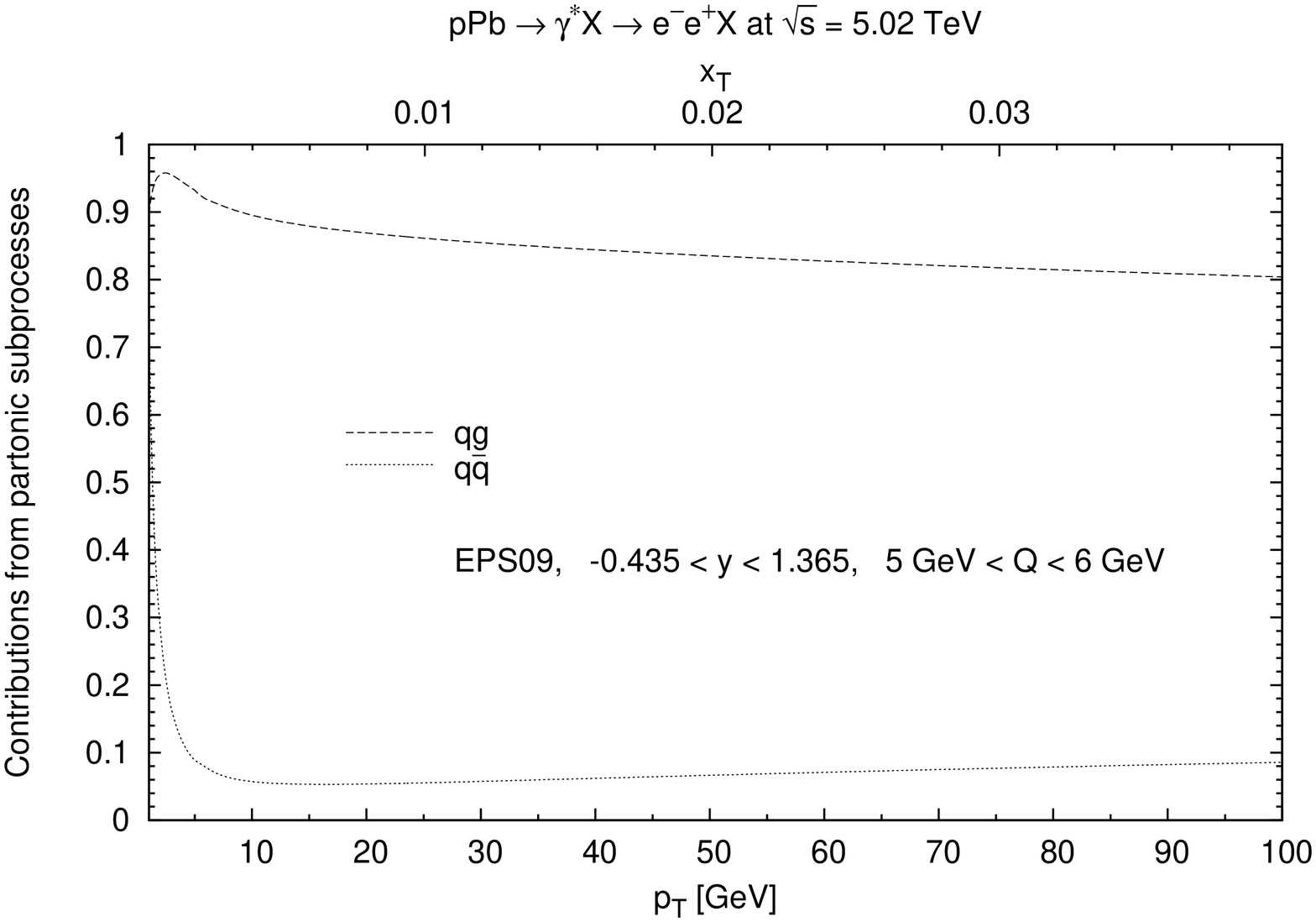,width=0.66\columnwidth,angle=-90}
 \caption{\label{fig:2}Fractional contributions of partonic subprocesses
 to low (top) and intermediate (bottom) mass electron-positron pair
 production at central rapidity and $p_T<100$ GeV.}
\end{figure}

In Fig.\ \ref{fig:2}, we show the NLO fractional contributions of
%
%
the quark-gluon (dashed) and quark-antiquark (dotted) processes to the
inclusive central production of electron-positron pairs with invariant
mass $2m_\mu<Q<2.5$ GeV  (top) and 5 GeV $<Q<6$ GeV (bottom)
in proton-lead collisions at the LHC with 
a centre-of-mass energy of $\sqrt{s}=5.02$ TeV. Processes like
$qq\to\gamma^*qq$ entering only at NLO are small and not shown.
For the nuclear PDFs, we have employed the central set of EPS09.
As it can be seen, the QCD Compton process takes over very quickly,
contributing more than 80\% (top) and even more than 95\% (bottom)
already at small, but finite values of $p_T$. 
From the upper $x$-axes of these plots, we learn that the
momentum fractions of the bound nucleons probed with $p_T<100$ GeV,
estimated by $x_T=2p_T/\sqrt{s}$, lie in the range 0.002 to 0.04,
i.e.\ in the shadowing and antishadowing regions (cf.\ Fig.\
\ref{fig:0}). The true $x$-values probed in the lead ion, travelling
here in the negative $z$-direction, are somewhat smaller ($x_{\rm Pb}/x_T=
e^{-0.465}\simeq0.63$) due to the asymmetry of the beam energies.

\begin{figure}
 \centering
 \epsfig{file=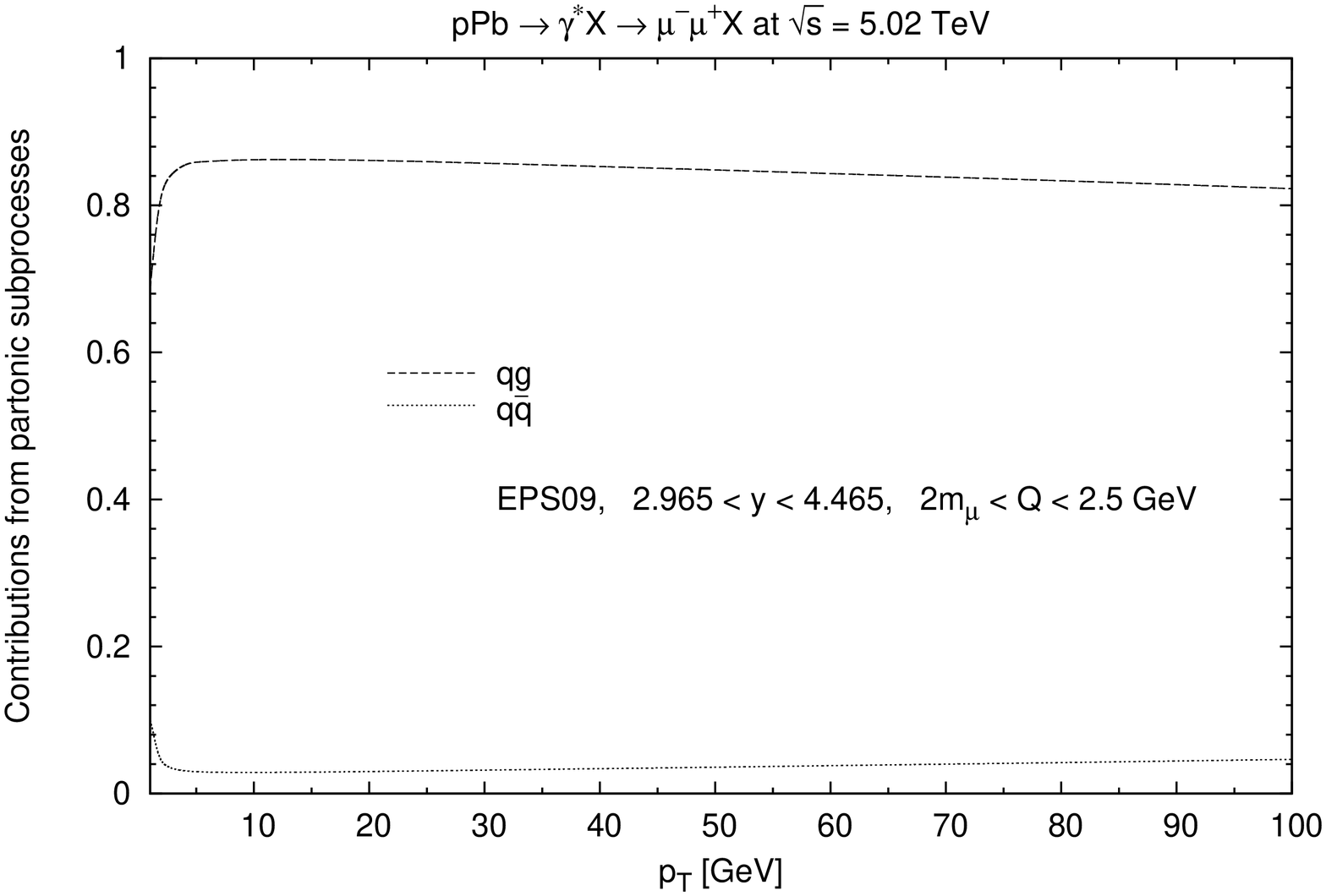,width=0.66\columnwidth,angle=-90}
 \epsfig{file=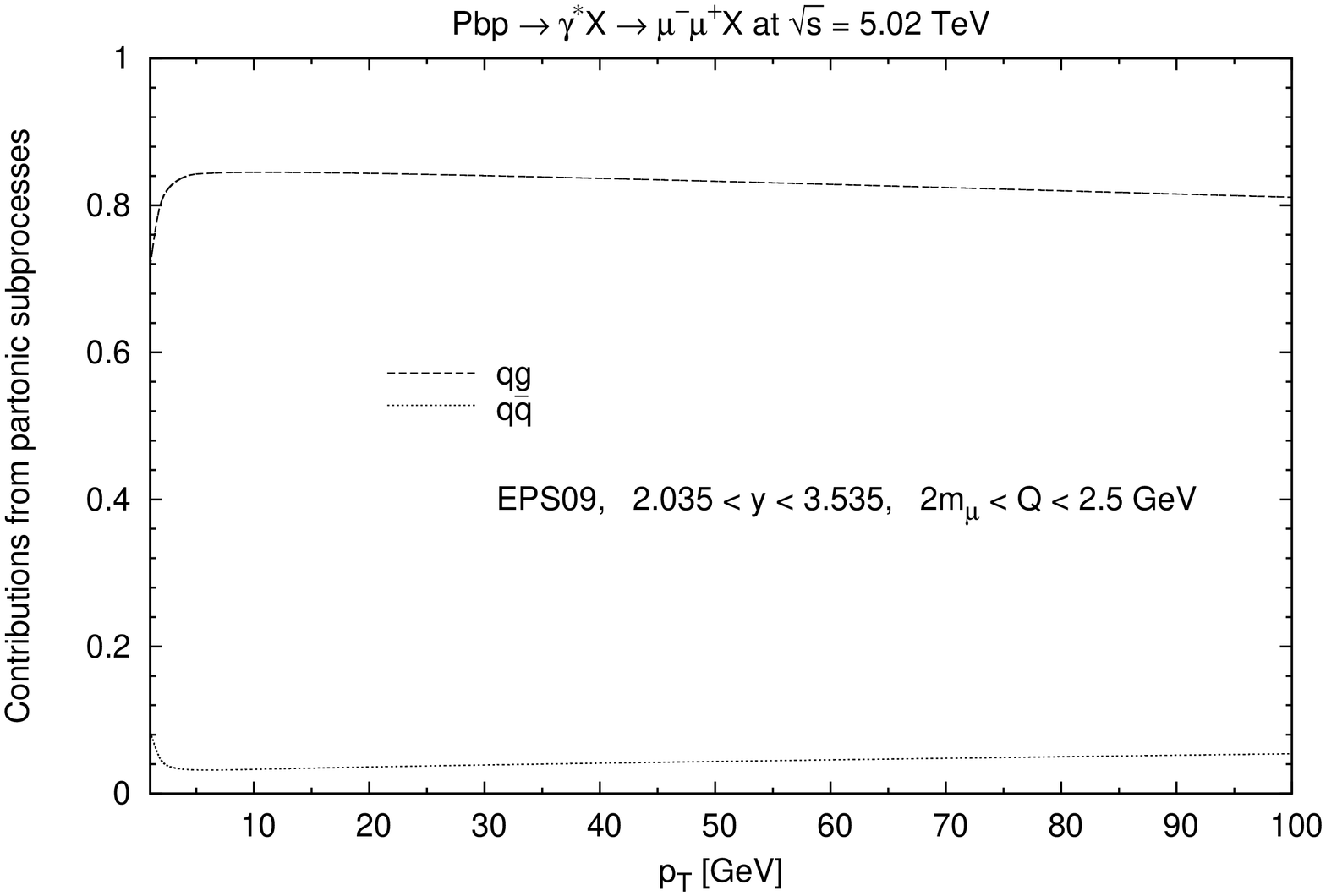,width=0.66\columnwidth,angle=-90}
 \caption{\label{fig:3}Fractional contributions of partonic subprocesses
 to low-mass muon pair production at forward rapidity and $p_T<100$ GeV
 in proton-lead (top) and lead-proton (bottom) collisions.}
\end{figure}

In Fig.\ \ref{fig:3}, similar results are shown for muons
%
%
produced in the forward region of the ALICE experiment. As in Fig.\
\ref{fig:2} (top), the virtuality of the photon lies just above the
threshold for muon pair production, and QCD Compton scattering
dominates again at the level of more than 80\%. However, the asymmetry
introduced by the acceptance of the muon spectrometer is now substantial,
so that the $x$-values probed in the lead ion depend strongly on
the orientation of the muon beam. Taking the central values of $x_T$
and muon acceptance, we find for lead ions circulating in the negative
$z$-direction, Fig.\ \ref{fig:3} (top), $x_{\rm Pb}\simeq0.02\cdot e^{-3.75}
\simeq4\cdot10^{-4}$, so that we are here firmly in the shadowing region.
For lead ions circulating in the positive $z$-direction, Fig.\ \ref{fig:3}
(bottom), the overall dominance of QCD Compton scattering changes very
little, since the asymmetry due to the beam energies is subdominant compared
to the one introduced by the ALICE muon spectrometer. However, we are now
probing mostly (valence) quarks in the lead ion with gluons from the proton,
so that isospin effects will become very important. Furthermore, we find
$x_{\rm Pb}\simeq 0.02\cdot e^{2.75}\simeq0.3$, so that we are here also in
the region of the EMC effect.

\begin{figure}
 \centering
 \epsfig{file=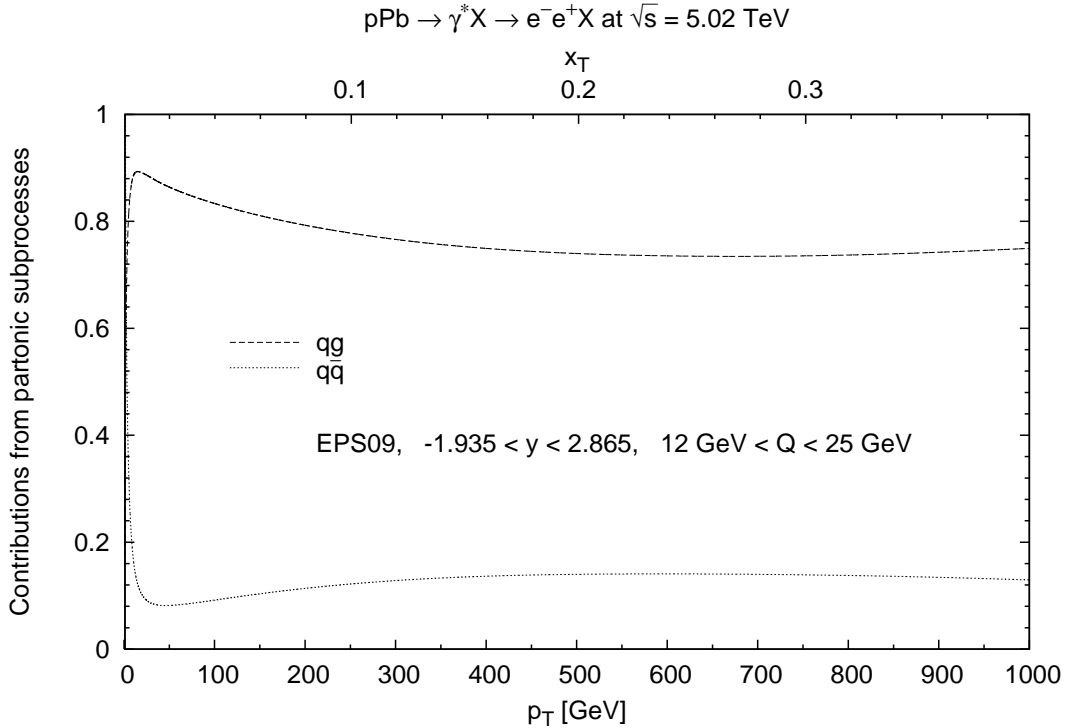,width=0.66\columnwidth,angle=-90}
 \caption{\label{fig:4}Fractional contributions of partonic subprocesses
 to intermediate-mass electron-positron pair production at central rapidity
 and $p_T<1000$ GeV.}
\end{figure}

In Fig.\ \ref{fig:4}, we study the fractional contributions of partonic
%
%
subprocesses in the ATLAS and CMS experiments with larger electron
and muon acceptance at central rapidity $|\eta|<2.4$ and a larger
transverse momentum range $p_T<1000$ GeV, but requiring also a larger
invariant mass of the lepton pair, e.g.\ 12 GeV $<Q<$ 25 GeV. For
these photon virtualities, the dominance of QCD Compton scattering
is a bit smaller than for the region between the $J/\Psi$ and $\Upsilon$
resonances shown in Fig.\ \ref{fig:2} (bottom), but still substantial
with more than 70\% in the complete $p_T$-range. With ATLAS and CMS,
larger $x_T$-values of 0.01 to 0.4 could be probed, where the true
$x_{\rm Pb}$ values depend in addition on the beam direction and possible
additional experimental rapidity cuts, but where we are mostly in
the regions of isospin and EMC effects.

\section{Low-mass lepton pair production at the LHC}
\label{sec:5}

In the previous section, we have established the dominance of the QCD
Compton scattering process over quark-antiquark annihilation for low-mass lepton
pair production in proton-lead collisions at the LHC. We have also
determined the kinematic regions where the ALICE, ATLAS and CMS experiments
are expected to be sensitive to shadowing, antishadowing, EMC and isospin
effects that modify the gluon and quark distributions of bound nucleons.

We can therefore now study numerically, at NLO QCD, the level of uncertainty
present in the current parametrisations of nuclear PDFs in the transverse
momentum distributions of lepton pairs in order to determine the potential
for their reduction with the corresponding measurements at the LHC. We will
do this through ratios of differential cross sections
\bea
 R^{\rm pPb}&=&{\d\sigma^{\rm pPb}/\d p_T\over \d\sigma^{\rm pp}/\d p_T},
\eea
where theoretical uncertainties, in particular those coming from variations
of the unphysical renormalization and bare-proton factorisation scales and
of the parametrisation of the bare-proton PDFs, are expected to cancel out
to a large extent. However, the nuclear modification factor may still show
some residual factorisation scale dependence. We stress again that our
numerical results have been obtained taking the full convolution of partonic
cross sections and PDFs in the numerator and denominator into account.

\begin{figure}
 \centering
 \epsfig{file=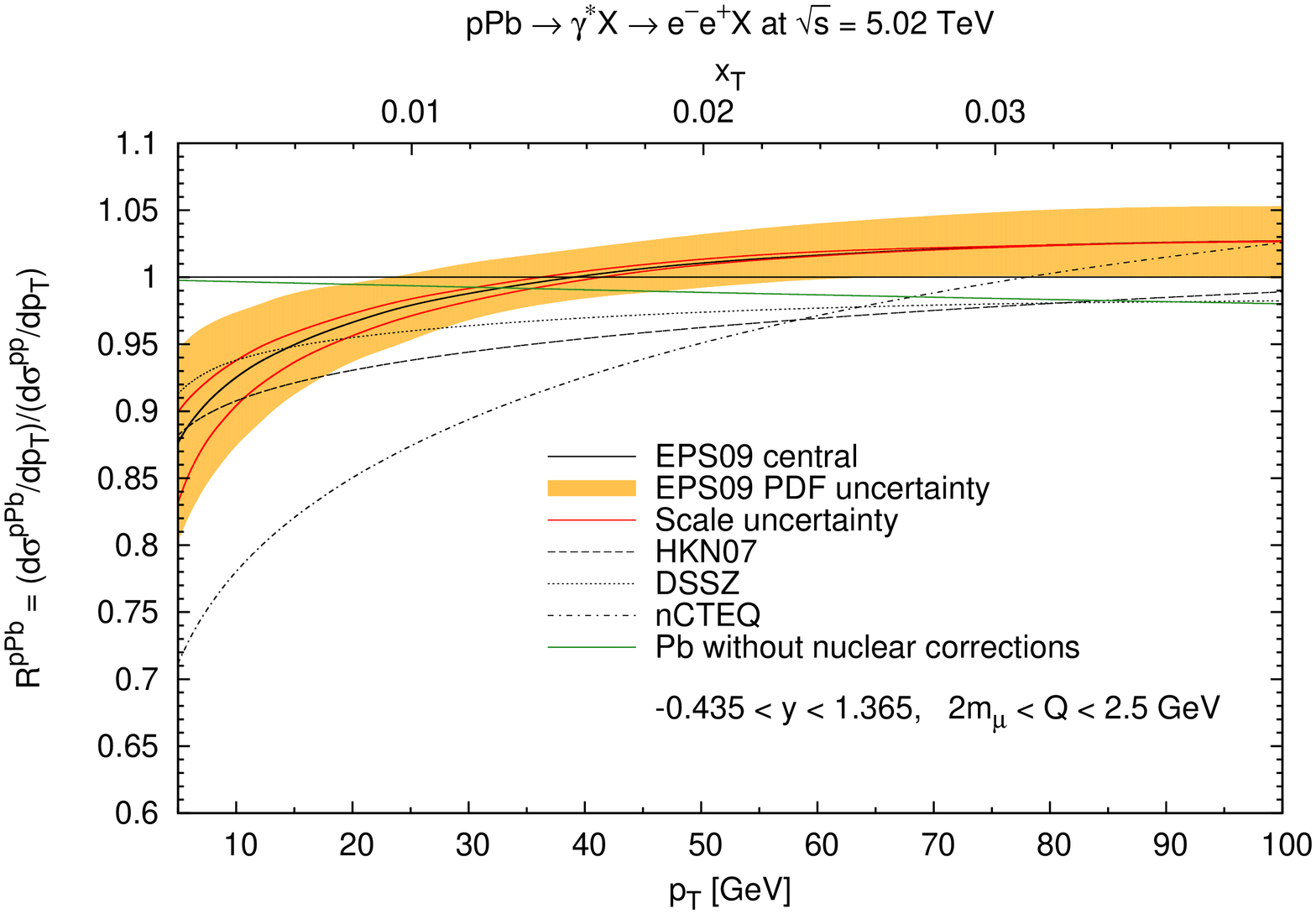,width=0.66\columnwidth,angle=-90}
 \epsfig{file=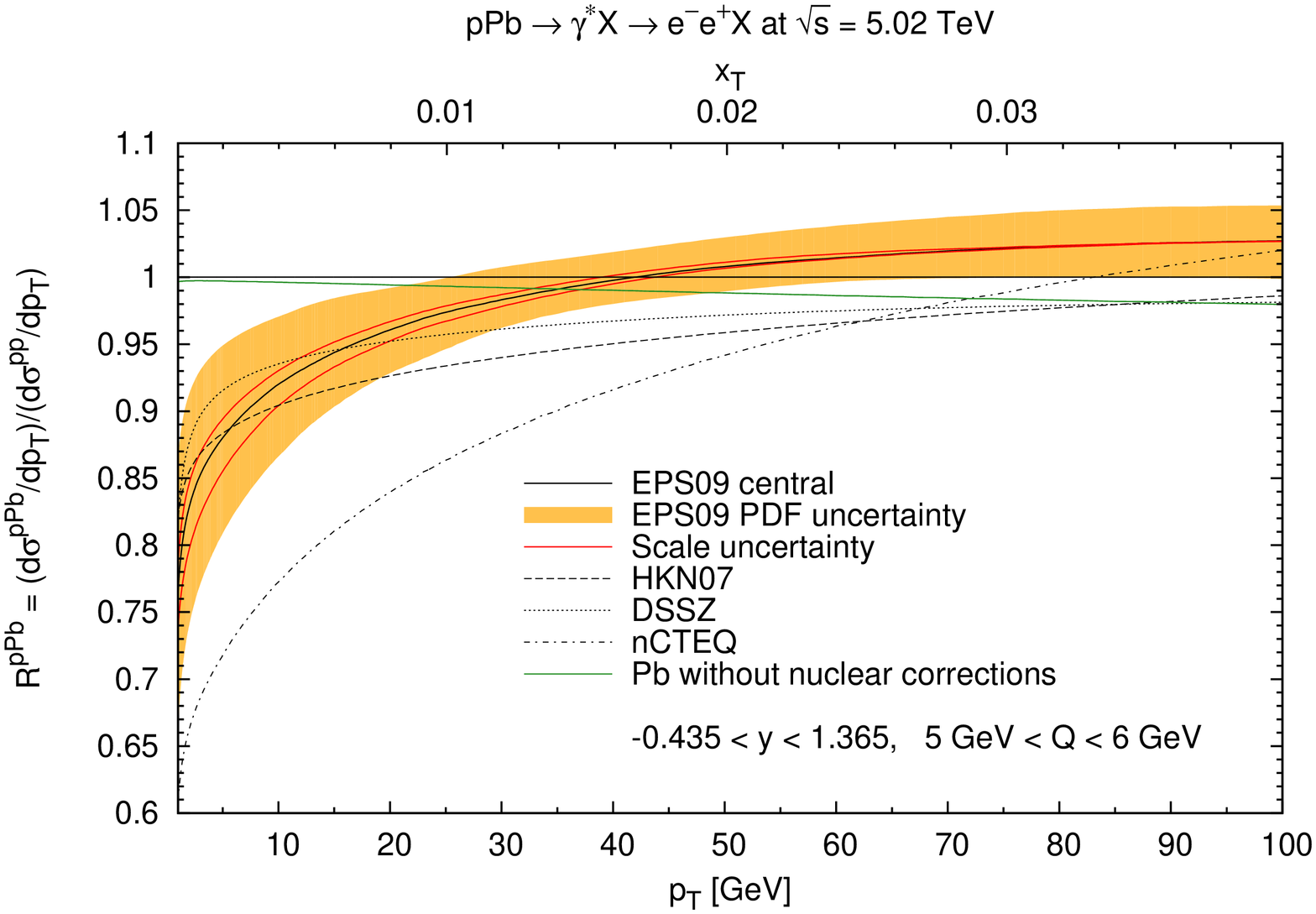,width=0.66\columnwidth,angle=-90}
 \caption{\label{fig:5}Ratios of transverse-momentum distributions of
 low (top) and intermediate (bottom) mass electron-positron pairs produced
 centrally in proton-lead and proton-proton collisions as predicted
 from various nuclear PDF parametrisations.
}
\end{figure}

In Fig.\ \ref{fig:5} (top), this ratio is shown for the acceptance
%
%
of the central ALICE detector and various nuclear PDF parametrisations.
Apart from the central EPS09 prediction (full), we show its associated
uncertainty band (yellow). As expected, the intermediate $x$-range from
0.002 to 0.04 corresponds to the transition from the shadowing to the
antishadowing regions, where the valence quarks are relatively well
determined (cf.\ Fig.\ \ref{fig:0}) and the overall uncertainty is
estimated by EPS09 to be relatively small, i.e.\ $\pm 6$\% at small
and 2.5\% at larger $p_T$. Luckily, the residual scale dependence 
(red), when the scales are varied simultaneously in the numerator
and denominator,
is even smaller by about a factor of two at small $p_T$, and it
is practically absent at larger $p_T$. However, one should not forget
the theoretical bias and restricted data set entering each global
nuclear PDF parametrisation. It is indeed instructive to see that
the older HKN07 parametrisation (dashed), but also the very recent
DSSZ parametrisations fall outside the EPS09 uncertainty band, differing
quantitatively by up to 8\% and qualitatively even about the nature
of the nuclear effects in this $x$-region. The discrepancy is even
more striking for the nCTEQ parametrisation, which is obtained without
a factorised ansatz and differs by up to 16\% from the central EPS09
prediction at low $p_T$. We also observe that isospin effects (green)
play no role here, since we are in a sea- and gluon-dominated regime.
The situation is very similar if we slightly increase in Fig.\ \ref{fig:5}
(bottom) the virtuality of the produced photon, so that it falls between
the $J/\Psi$ and $\Upsilon$ resonances.

With the ALICE muon spectrometer at forward rapidities, substantially
%
\begin{figure}
 \centering
 \epsfig{file=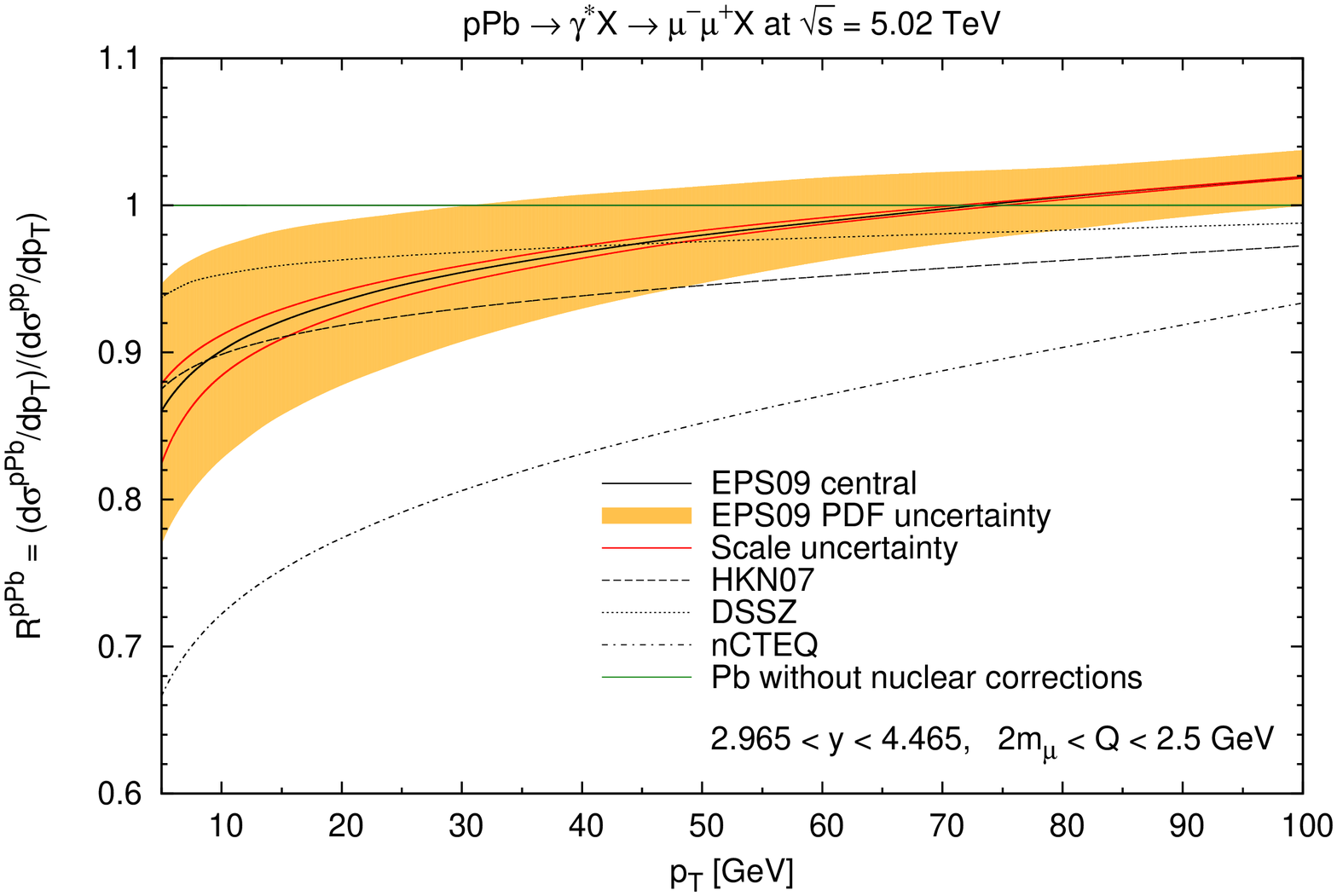,width=0.66\columnwidth,angle=-90}
 \epsfig{file=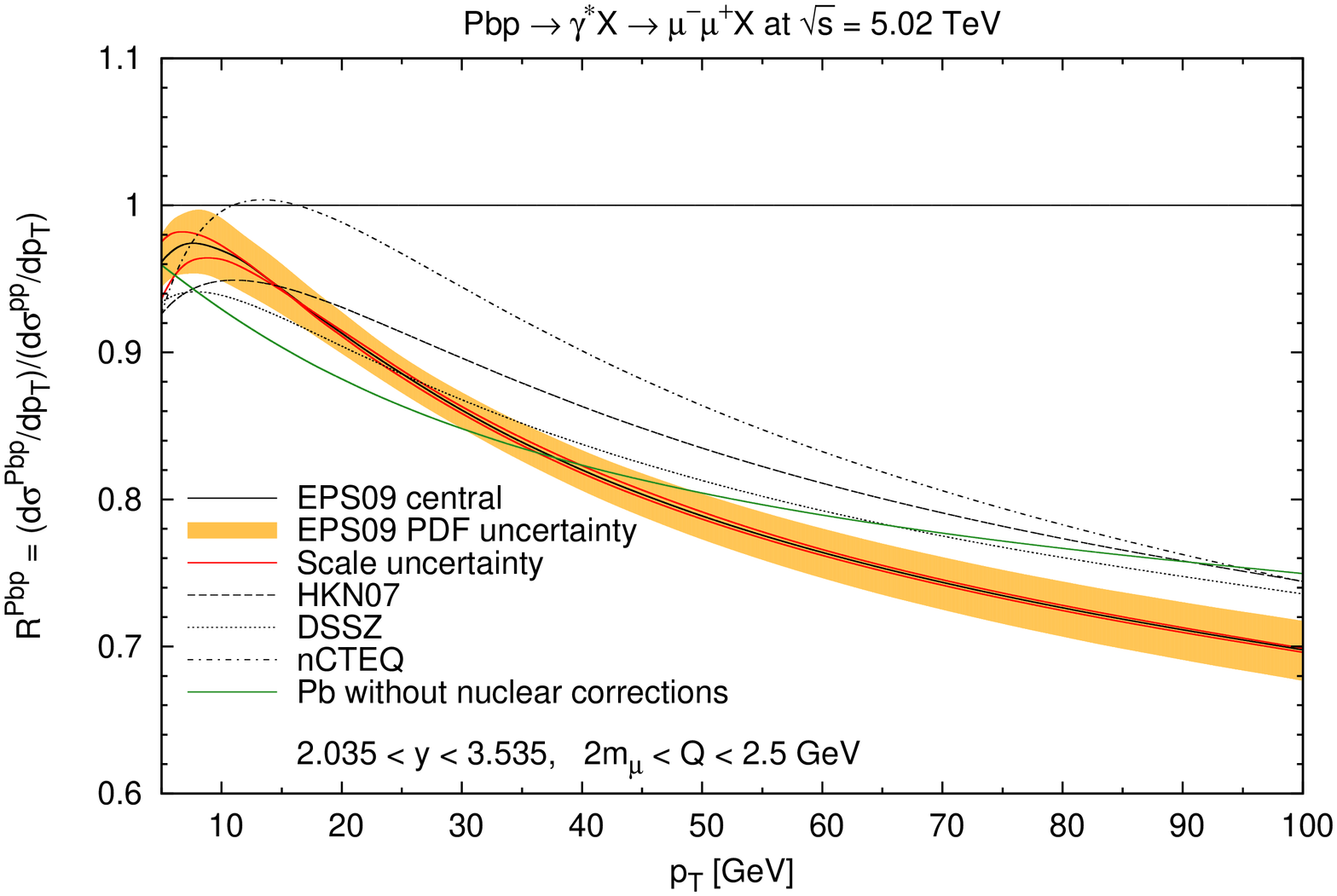,width=0.66\columnwidth,angle=-90}
 \caption{\label{fig:6}Ratios of transverse-momentum distributions of
 forward muon pairs produced in proton-lead (top) / lead-proton (bottom)
 and proton-proton collisions as predicted from various nuclear PDF
 parametrisations.}
\end{figure}
%
smaller $x$-values in the bound nucleons are probed, so that we are
moving more into the shadowing region. This is confirmed by Fig.\ \ref{fig:6}
(top). The EPS09 uncertainty now reaches $\pm 9$\% at low $p_T$.
The residual scale uncertainty there is again much smaller (up to 2\%),
so that LHC measurements could be reliably used to constrain the nuclear
PDFs. nCTEQ predict again much stronger shadowing of more than 32\%,
contrary to 14\% for EPS09, 12\% for HKN07 and even only 6\% for DSSZ.

If the proton and lead beam directions are reversed, the situation changes
dramatically, as can be seen from Fig.\ \ref{fig:6} (bottom). The ALICE
forward muon spectrometer is now mostly probing bound valence quarks, so that
all parametrisations are following to first approximation the prediction
from isospin effects only (green). Furthermore, the central EPS09 curve predicts
antishadowing up to $p_T$ of about 40 GeV and an EMC effect beyond this value.
The other parametrisations see this transition only at the high end of the
$p_T$-range shown here, i.e.\ at 70, 90 and 95 GeV for DSSZ, HKN07 and nCTEQ,
respectively. They fall outside the EPS09 uncertainty band, which is quite small
($\pm 2$\%), over almost the entire $p_T$-range, differing from it by up to 8\%.
Scale uncertainties are again very small, so that the LHC measurements
proposed here can be of great importance not only in the small-, but also in the
large-$x$ regions. 

With the central detectors of ATLAS and CMS, it should be possible to
%
\begin{figure}
 \centering
 \epsfig{file=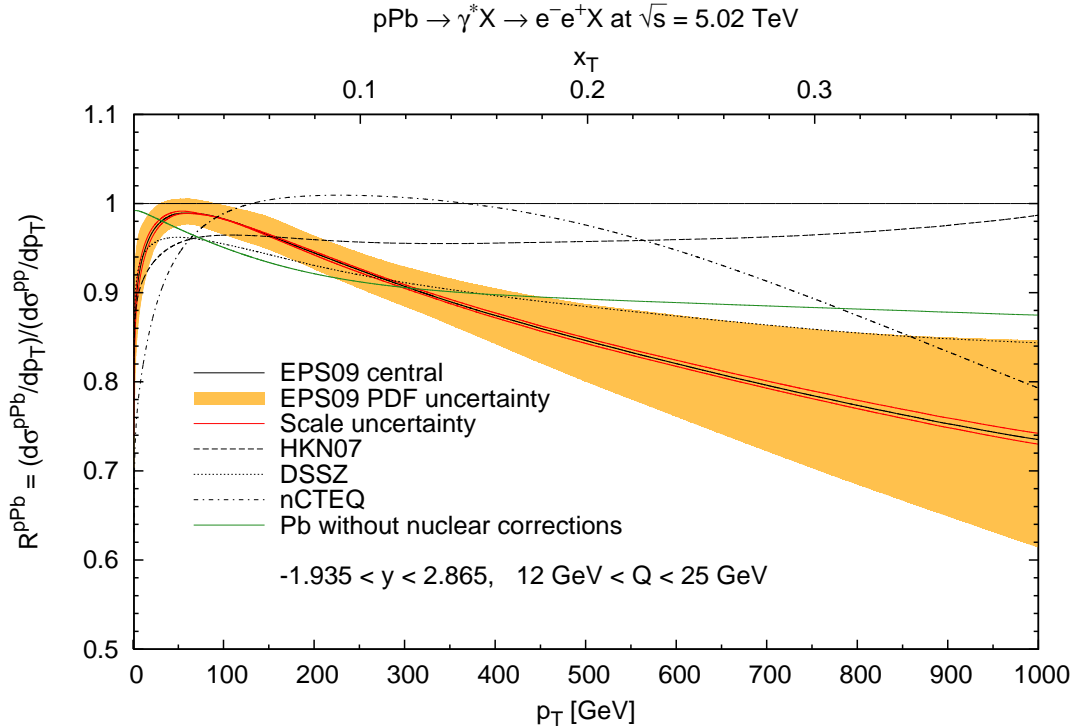,width=0.66\columnwidth,angle=-90}
 \caption{\label{fig:7}Same as Fig.\ \ref{fig:5} for slightly larger
 dilepton invariant mass and considerably larger transverse momenta.}
\end{figure}
%
measure electron-positron and muon pairs with slightly larger invariant mass and
considerably larger transverse momenta than in the ALICE experiment, but
nevertheless providing again access to the regions with antishadowing,
isospin and EMC effects as with the ALICE forward muon spectrometer.
This is confirmed by Fig.\ \ref{fig:7}, which is qualitatively similar
to the Fig.\ \ref{fig:6} (bottom). Isospin effects soon take the
leading role. According to the EPS09 and DSSZ predictions, the transition from
antishadowing to the EMC effect should happen at 300 GeV. Below (above)
this value, the DSSZ prediction has larger antishadowing and smaller EMC
effects and lies close to or at the edge of the EPS09 uncertainty band.
The shapes of HKN07 and nCTEQ are very different, which can be attributed
to smaller/older data sets and a more general parametrisation of nuclear
effects, respectively. With nCTEQ in particular, the transition to the
EMC effect region occurs only at very large $p_T$. A similar behavior had
already been observed in Fig.\ \ref{fig:6} (bottom).

\section{Conclusions}
\label{sec:6}

In conclusion, we have proposed in this paper measurements of low-mass
lepton pair production in proton-lead collisions at the LHC to better
constrain the nuclear parton densities. After
reminding the reader about our currently very unsatisfactory knowledge
of nuclear corrections, in particular for the gluon at small $x$, but
also for other partons at large $x$, we established the dominance of
quark-gluon scattering in analogy with real photon production, but
without its theoretical and experimental fragmentation and isolation
uncertainties. Using NLO QCD calculations and current experimental
acceptances of electrons and muons in the ALICE, ATLAS and CMS
experiments, we numerically demonstrated that measuring ratios of
transverse momentum distributions has a large potential to reduce
the theoretical uncertainties on the nuclear modifications of PDFs
or even rule out some parametrisations.
Depending on beam orientation and central or forward kinematics,
all interesting regions, shadowing, antishadowing, isospin and effects
can be probed, while scale and bare-proton PDF uncertainties,
as well as many systematic experimental errors, should cancel out.
The results of these measurements could subsequently be used to
establish new, more reliable nuclear PDFs in global fits.

\acknowledgments

We thank C.\ Klein-B\"osing for useful discussions and comments on the
manuscript.




\end{document}